\documentclass[preprint]{aastex}
\usepackage{psfig}

\newcommand{\etal}{\emph{et al.\ }}
\newcommand{\qmin}{$Q_{\rm min}\ $}

\newcommand{\tbin}{$T_{\rm Bin}$}
\newcommand{\irsfive}{L1551~IRS~5}
\newcommand{\lsun}{$L_\odot$}
\newcommand{\msun}{$M_\odot$}

\begin{document}

\title{Planet formation is unlikely in equal mass binary
systems with $a\sim50$~AU} 

\author{Andrew F. Nelson}
\affil{Max Planck Institute for Astronomy, K\"onigstuhl 17, 
D-69117 Heidelberg, Germany}

\begin{abstract}

We show that planet formation via both gravitational collapse and core
accretion is unlikely to occur in equal mass binary systems with
moderate ($\sim 50$~AU) semi-major axes. Internal thermal energy
generation in the disks is sufficient to heat the gas everywhere so
that spiral structures quickly decay rather than grow or fragment.
This same heating will inhibit dust coagulation because the
temperatures rise above the vaporization temperatures of many volatile
materials. We consider other processes not included in the model and
conclude that our temperatures are conservatively estimated (low), i.e.
planet formation is less likely in real systems than in the model.

\end{abstract}


\section{Introduction and the Model Specification}

Both indirect evidence \citep{ALS88,BSCG} and later direct imaging
\citep{Close97,MO96}, have shown that disks are quite
common in young stellar systems. These disks are commonly thought
\citep{BS96} to be sites for planet or brown dwarf formation. A large
fraction of stars are formed in binary systems \citep{MGJS_PP4} and in
the same star formation regions as single stars. Theory suggests that
the most likely mechanisms responsible for forming Jovian mass planets
or low mass brown dwarfs are either gravitational collapse of large
scale spiral structure or coagulation of small solid grains followed
by later accretion of additional gas (`core accretion') in the disks
of forming stellar systems. Evaluating the effectiveness of these
mechanisms is important for understanding the origin of our own
solar system as well as planetary systems in other mature single or
multiple systems.

The \irsfive\ system serves as a useful observational testbed for
comparison to theoretical modeling because of its relative youth
\citep[$\sim10^5$~yr,][]{BTC94} and many previous detailed
observations \citep[see e.g.][]{Men-Hen-97}. This system consists of
an extended nebulosity some 2400$\times$1100~AU in size with an inner
core of 220$\times$76~AU \citep{Mom98}. Two bipolar jets flow outward
in each direction from the core in the plane perpendicular to its long
axis. The core has been resolved into two sources with projected
separation of about 50~AU and inferred disk masses of $\sim.05$\msun,
each $\sim$20-25~AU in diameter \citep{Rodriguez98}. The total mass in
the core has been estimated to contain 0.5--1.0\msun\ of material
\citep{ALS88,Mom98}, which produces $\sim$30\lsun\ in luminous output
\citep{KeeMas90}. 

We present a numerical simulation of a binary star/disk$+$star/disk
system using a two dimensional ($x,y$) Smoothed Particle Hydrodynamic
(SPH) code. The dimensions of the disks and semi-major axis of the
binary are chosen to be similar to the inner core region of \irsfive.
In the absence of strong constraints on the constituents of the binary
(e.g. the masses of the two stars), we choose to set up a binary system
consisting of identical components, obtained by setting up a single
system in isolation, then duplicating it exactly. We assume each star
and disk have mass $M_*=0.5$\msun\ and $M_D=0.05$\msun, respectively. 
The disk radius is set to $R_D=15$~AU which, for a semi-major axis of
$a=50$~AU, is comparable to the largest stable streamline
\citep{Pac77}.

The mass and temperature of the disk are distributed according to
$r^{-3/2}$ and $r^{-1/2}$ power laws respectively. The absolute scale
of each power law is determined from the disk mass, the radial
dimensions of the disk and the condition that the Toomre stability
parameter, $Q$, is no smaller than \qmin=1.5 over the entire disk.
This value ensures that the simulation begins in a state marginally
stable against the growth of spiral structure, so that we do not
accidently `discover' a collapsed object early in the evolution which
in reality is an artifact of our initial condition. Both density and
temperature are free to vary in time and space, so the initial
condition will not prevent spiral structure growth or fragmentation,
if the evolution leads to such. The gas is set up on circular orbits
around the star so that pressure and gravitational forces exactly
balance centrifugal forces. Radial motion is zero. The magnitudes of
the pressure and self-gravitational forces are small compared to the
stellar gravity, so the disk is nearly Keplerian in character.

Approximately 60000 equal mass particles are set on a series of
concentric rings around the star in a single, star/disk system, then
duplicated, bringing the total number of particles to $\sim$120000.
The two stars and disks are offset equal distances in the $+x$ and
$-x$ directions. We define the binary semi-major axis to be $a=50$~AU,
similar to \irsfive. Only weak constraints on eccentricity exist in
\irsfive, primarily consisting of the sizes of the observed disks:
eccentricities larger than $e=0.3$ would lead to rapid Roche lobe
overflow. We set $e=0.3$ to be the initial value in this simulation.
The system is at apoapse at time $t=0$ with the orbital velocities
defined by approximating each star$+$disk system as a point mass, so
that the orbit determination reduces to the solution of the two body
problem.

The disks are self gravitating and each star is modeled as a point
mass free to move in response to gravitational forces from the rest of
the system. The stellar gravitational forces are calculated using a
Plummer potential with a softening radius of 0.2~AU, which also serves
as an accretion radius, $r_{acc}$. SPH particles with trajectories
that pass closer than $r_{acc}$ to a star are absorbed, and the star's
mass and momentum increase accordingly.

The thermodynamic evolution is identical to that described in
\citet{DynamII}. Thermal energy is added to the gas due to active
hydrodynamic processes using an artificial viscosity scheme, which
approximately models shocks and turbulence. This heating is roughly
equivalent in magnitude to an alpha model with $\alpha\sim 2-5
\times10^{-3}$. Thermal energy is removed from the disk gas by
radiative cooling due to passive blackbody emission from the disk's
photosphere surfaces. The blackbody temperature is calculated at each
time step and for each SPH particle. This treatment remedies a major
shortcoming of previous models \citep{DynamI,Pick98,Boss97} which used
a `locally isothermal' or `locally adiabatic' approximation to show
that relatively low mass disks can undergo fragmentation and/or
collapse, despite earlier claims \citep{PodPP3} that a very massive
disk is required.

\section{Results}

\begin{figure} \begin{center}
\psfig{file=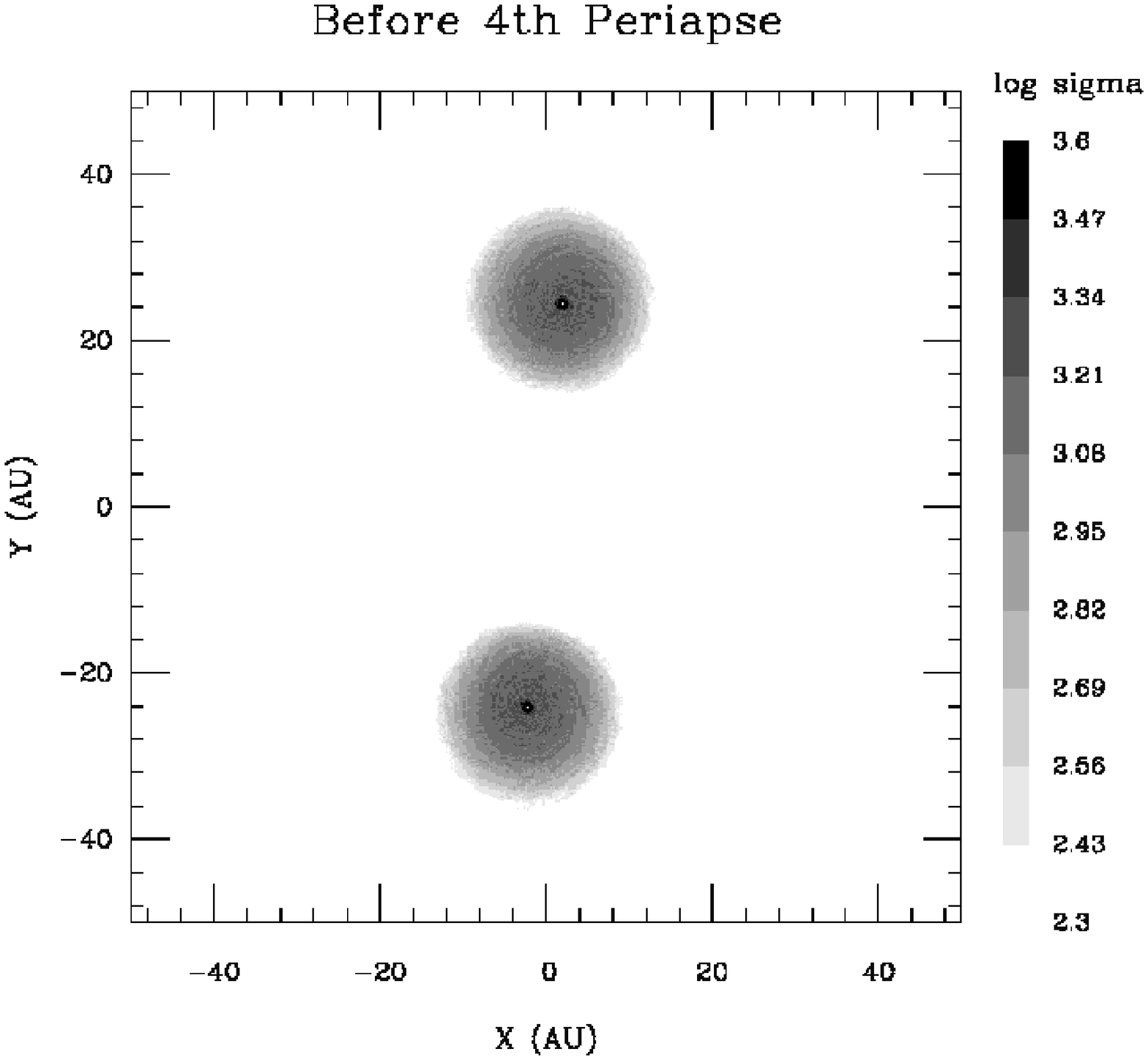,angle=-90,height=100mm}
\psfig{file=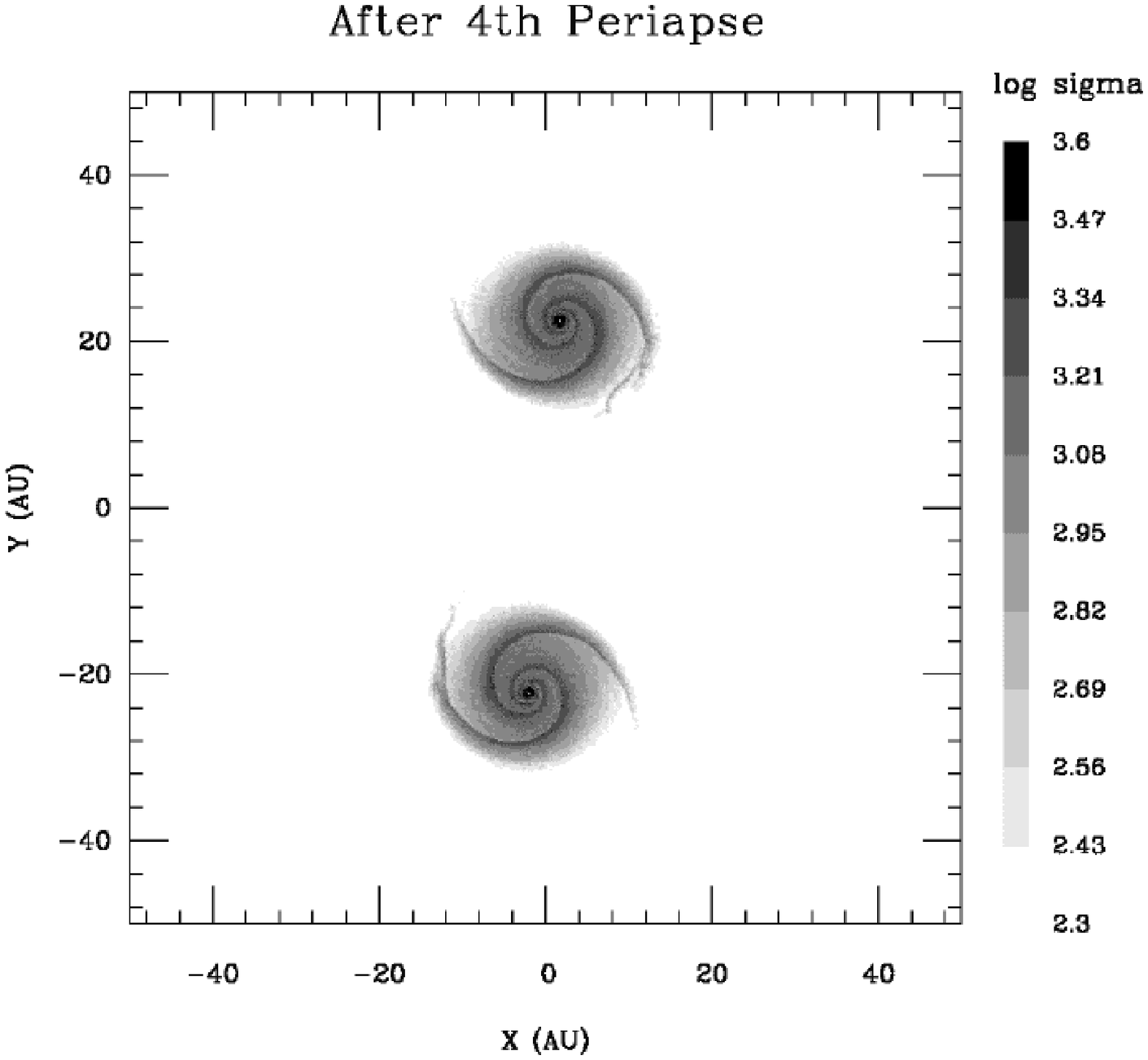,angle=0,height=100mm,rheight=90mm}
\end{center}
\caption
{\label{fig:disk-pair} 
The particle distribution of the binary system before (top) and after
(bottom) periapse passage. Mass surface density units are in
$\log_{10}({\rm gm/cm}^2)$. The trajectory of each component is
counterclockwise and periapse occurs when the stars (at each disk
center) reach the $y=0$ axis and are 35~AU apart. No structure is
visible in either disk, except that they are no longer exactly round.
In the bottom panel, the two components have reversed positions from
that shown in the top panel. Tidal torques have caused two armed
spiral structures to develop in the disks. In both images, these
torques have also caused mass to be redistributed. The disk edge is no
longer sharp and is found near an average radius of $\sim$12-13~AU
rather than the initial 15~AU.} 

\end{figure}

The system is evolved for eight binary orbits, or 2700~yr
(\tbin=350~yr). Figure \ref{fig:disk-pair} shows the system shortly
before and after the fourth periapse passage of the two components
(measured from the beginning of the simulation). Before each periapse,
the two disks are smooth and exhibit no visible spiral structure,
although they are no longer perfectly `round'. During and after
periapse, each disk develops strong, two armed spiral structures due
to the mutual tidal interactions of the binary. The structures decay
to a smooth condition like that in the top panel over the next
$0.5$\tbin. The cycle repeats with little variation as the system
again approaches periapse, and we expect that further evolution will
be similar.

\begin{figure}
\begin{center}
\psfig{file=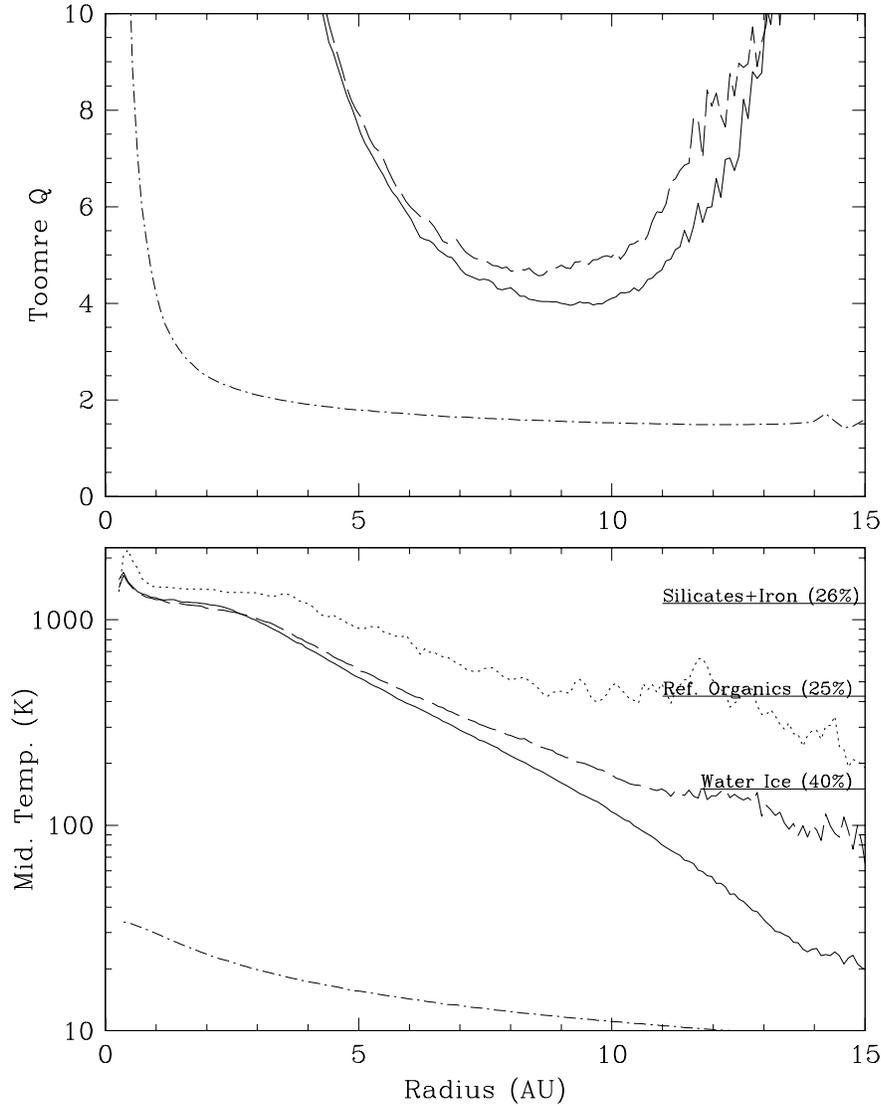,height=160mm}
\end{center}
\caption{\label{fig:temp-prof} 
The azimuth averaged Toomre $Q$ (top) and temperature (bottom)
profiles of the disks shown for the same times before (solid) and
after (dashed) periapse as in figure \ref{fig:disk-pair}. The initial
profiles are shown with dashed dotted lines. Both show large increases
over their initial values at all radii. The large increase in $Q$
outside 12~AU is due to the truncation of the disk and relative
scarcity of matter remaining there. The dotted curve shows the maximum
temperature reached inside the spiral arms at that radius. At the
right are vaporization temperatures of the major grain species in the
solar nebula and their fraction of the total grain mass, as discussed
in \citet{Pol94}. }

\end{figure}

The spiral structures decay because internal heating in the disk
increases the stability of the disks against spiral arm growth, as
measured by the Toomre stability (fig. \ref{fig:temp-prof} top) of
each disk. The minimum value of $Q$ increases from its initial
\qmin=1.5 to \qmin$\sim4$ before periapse and \qmin$\sim5$ afterwards.
These values are the same before and after each successive periapse
passage and both are well above the $Q\lesssim3$ values for which
spiral structures are expected to grow. Therefore, if we suppose that
Jovian planets form via gravitational collapse or fragmentation of
spiral structure in disks, their formation will be unlikely in this
system. 

The high stability is due to an increase in the disk temperature. In
fact, the temperatures are high enough to cause some grain species to
be vaporized. This is important because the core accretion model for
planet formation requires that solid grains can coagulate and are not
instead repeatedly returned to vapor state. Water ice may be
particularly important because it composes 40\% by mass of the solid
material in the disk and is among the most volatile grain species
\citep{Pol94}, vaporizing at $\sim150$~K. 

The bottom panel of figure \ref{fig:temp-prof} shows the disk midplane
temperatures obtained in this simulation. Only in a region with
$r\gtrsim 10$~AU does the temperature reach low enough values that
water ice can form, even temporarily. However in this region, the
matter is most subject to shocks generated by the spiral structure
produced by the binary interaction, which raise its temperature by as
much as a factor three over the azimuth average. The spiral patterns
co-rotate with the orbit of the binary and the orbital period at 15~AU
is $\sim$82~yr, so material everywhere in the disk will have time to
travel through the spiral arms several times before they decay. 
A simulation run with zero eccentricity also produces spiral structure.
In this run the azimuth averaged temperatures were similar to fig.
\ref{fig:temp-prof}, but the temperature in the shocked regions are
not as extreme. Water ice would still be vaporized everywhere but more
refractory species may not be.

In the outer disk, grains less than $\sim 1$~mm in size which pass
through such a shock and which contain water ice will be vaporized on
a timescale of $\lesssim10^4-10^5$~s depending on the temperature
\citep{ELL90,LGH95} and the remaining more inert species may
dissagregate. In this region, passage through the warmest part of
a spiral arm requires $\sim 1-2$~yr, so sufficient time exists to
return grains of this size to the gas phase. Grain growth may still
occur between the spiral arms and when the spiral arms have decayed,
but must begin with gaseous material each time, so growth of solid
material into larger entities will be suppressed. Temperatures at high
altitudes are lower, but grains which form there will tend to sink to
the midplane as they grow larger and also be destroyed. Therefore
Jovian planet formation by the core accretion mechanism will occur
much more slowly, if at all in this system.

The weak link remaining in the argument against the core accretion
mechanism is the lack of knowledge of the microphysics important for
dust coagulation. For example, one could imagine that growth of 
silicate and iron grains is catalyzed by temporarily enhanced cross
sections as mantles of more volatile material form on their surfaces
and the gravitational torques produced by the binary interaction
enhance mixing throughout the disks. Even if this type of interaction
takes place, the eventual formation of planet sized objects remains in
doubt. As the rocky aggregates grow, conditions appropriate for dust
coagulation (e.g. `perfect sticking') break down and collisions
between particles become increasingly disruptive due to the finite
strength of the aggregates.

The disruption of solid bodies depends strongly on the relative
velocity of the impactor and target particles, with disruption
occurring for velocities $\gtrsim$1-3~km/s for planetesimal sized
targets ($\lesssim 1$~km) \citep{BA99}. On average in an accretion
disk, the relative velocity of planetesimals will be proportional to
their eccentricities, $v_{\rm rel}\approx ev_{\rm orb}$. For our
model, a relative velocity of 1~km/s corresponds to an eccentricity of
$e\sim 0.05$ at 1~AU, or $e\sim 0.15$ at 10~AU. 

We have seen that gravitational torques are strong enough to generate
large amplitude spiral structure as they drive the gas onto eccentric
orbits. Gas eccentricities are quickly damped due to shock
dissipation, but planetesimals are only weakly coupled to the gas and
will have time to encounter other objects and collide or to increase
their eccentricities still further as the evolution proceeds. If
particle eccentricities can grow to $e\sim0.1$, we expect that the
growth of kilometer sized bodies will be suppressed in binary systems
such as the one modeled here. However, a more detailed analysis must
be done in order to constrain this possibility.

\subsection{Checks on the validity of the conclusions}

The conclusions regarding planet formation will remain valid as long
as the temperatures determined from the model are lower limits on the
temperatures present in real systems. If the model produces
temperatures which are too low, then real systems will be even more
stable against spiral structure growth and fragmentation and less
likely to produce large, coagulated grains. If they are too high, the
model could inaccurately portray the disk as too stable. Are the
temperatures produced in the model too low? We can constrain the
temperatures in the model by comparing the radiated energy from
observed systems (here specifically to \irsfive), to that produced by
the simulation in various wavelength bands. This comparison requires
that we relate the luminous output to the temperature. 

In regions where the optical depth is high, as it is in the accretion
disks, the radiated emission can be approximated as a blackbody with a
temperature of the disk photosphere. The disk's midplane and
photosphere temperatures are then related to each other by a given
Rosseland opacity and the local vertical density/temperature profile.
We determine such a profile as by-product of the cooling model in this
work, under the assumption that the vertical structure is
instantaneously adiabatic. Other work \citep{BCKH,dall98} has shown
that the structure may instead be super adiabatic. If this is the case
then the midplane temperatures will be higher, and our conclusion is
stronger.

The opposite case may be true instead: large relative heating can
occur at high altitudes, even though the high altitude heating is
small in an absolute sense \citep{Pick00}. This means that vertical
temperature structure may become distorted and an incorrect midplane
temperature could be inferred. High altitude dynamical heating will
play a role similar to high altitude passive heating from stellar
photons. \citet{dall98} show that this process produces a temperature
inversion high above the photosphere, but this region contributes
negligibly to the radiated flux. Therefore, we can rely on the modeled
radiated output of the simulation to represent accurately the
temperatures at the disk midplane, given the physical processes
included in the calculation. Our conclusions about the formation of
planets will be confirmed if we find that the energy output from the
disks is equal to or less than that observed.

To obtain a valid comparison between the observed and modeled fluxes,
we must be certain that the flux from other parts of the system (e.g.
the circumbinary disk and envelope) is not a significant contributor
to the observed flux used in the comparison. We must also be certain
that extinction between the source and the observer has not altered
the emitted flux. We therefore require very high spatial resolution
photometry at long wavelengths, not affected by extinction. 

\begin{figure}
\begin{center}
\psfig{file=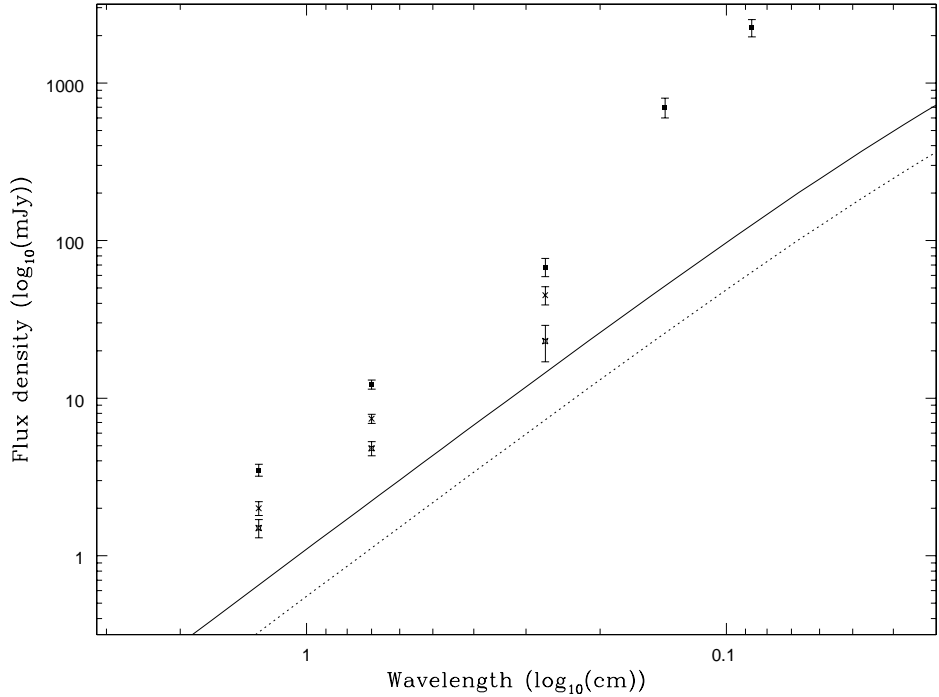,angle=-90,height=100mm}
\end{center}
\caption{\label{fig:freq-cmp} 
Long wavelength flux densities from \irsfive, the total from both
components of the binary are shown with solid symbols, and the
individual components are shown with open symbols. The model is
displayed with a solid line for the total from both components,
while the value for one component is shown with a dotted line. Each
assume a distance of 140~pc to the source. At each wavelength, the
model underestimates the flux, indicating that the model temperatures
are lower limits. The data are taken from the following literature at
an angular resolution given in parentheses: D. Koerner, (personal
communication, $\sim0.21$\arcsec at 1.3~cm), \citet{Rodriguez98}
($\sim0.06$\arcsec at 7~mm), \citet{LMW97} ($\sim0.3$\arcsec at
2.7~mm), \citet{woody} ($\sim3$\arcsec, at 1.4~mm) and \citet{LCHP94}
($\sim .8$\arcsec using single baseline interferometry at 870$\mu$m).
For comparison, the binary separation of the \irsfive\ system quoted
by \citet{Rodriguez98} is $\sim0.3$\arcsec. }

\end{figure}

In figure \ref{fig:freq-cmp}, the highest available resolution, long
wavelength observations of \irsfive\ are plotted and compared to the
flux densities produced from the simulation. For all wavelengths
between 1.3~cm and 870~$\mu$m, the observed fluxes exceed those
obtained from the simulation by a factor of $\sim 5$. The differences
at 1.4~mm and 870~$\mu$m are larger, a factor $\sim10$, however they
do not resolve the binary and may contain some contaminating flux
from the circumbinary environment. Given these comparisons, we can
conclude that the temperatures in the disk are conservatively
estimated by the simulation (too low) and that the disks in \irsfive\
are more stable against spiral arm growth and fragmentation and less
likely to allow dust coagulation than in the simulation. Our
conclusion that planet formation is unlikely in the \irsfive\ system
is secure. 

\section{Remarks and remaining questions}

The inventory of physical processes considered in this calculation is
not complete. The total luminosity will include direct contributions
at the short wavelength end of the spectrum from not only the two
circumstellar disks, but also the two stars. Accretion of material
from the inner disk edge (0.2~AU for this simulation) onto the stellar
surface will also add to the total. At longer wavelengths, the
circumbinary disk, the infalling envelope and radio emission from
bipolar polar outflows will add to the total. Of these processes, only
outflows remove thermal energy from the disks, but such outflows are
thought to originate very close to the star. They will not remove
energy from the part of the disk important for planet formation. The
rest will either add directly to the short wavelength spectrum or heat
the cooler parts of the disk by absorption of short wavelength
radiation, which subsequently reradiates at longer wavelengths
\citep[see e.g.][]{Bell99}. Including them in the calculation can only
raise disk temperatures and strengthen our conclusions. Estimates of
the total luminosity (not shown) from all these sources are within a
factor of two of the observed 30\lsun\ luminosity of \irsfive. We are
encouraged by this agreement, and expect that the overall agreement
between model and observations would become closer if radiative
reprocessing were included.

A number of questions remain. What is the distribution of planetesimal
random velocities? Will lower mass disks be more susceptible to planet
formation because of increased efficiency of radiative energy losses?
What happens for binary systems with different separations or with
unequal stellar mass ratios? How distant is distant enough, so that
one component of the binary does not strongly influence the other?
These questions will be addressed more completely in a followup paper.

\acknowledgements

I gratefully acknowledge David Koerner for the generous release of the
1.3~cm fluxes before their publication. I thank Brian Pickett and
Willy Kley for helpful conversations during the development of this
paper and for Brian's later comments as referee.

\singlespace

\clearpage

\clearpage

\end{document}